# Automated, Consistent, and Even-handed Selection of Active Orbital Spaces for Quantum Embedding

*Elena Kolodzeiski* and *Christopher J. Stein*

Technical University of Munich, TUM School of Natural Sciences, Department of Chemistry, Lichtenbergstr. 4, D-85748 Garching, Germany

Keywords: Subsystem DFT, Quantum Embedding, Active Space, Metal Nanoclusters, Heterogeneous Catalysis

Abstract: A widely used strategy to reduce the computational cost in quantum-chemical calculations is to partition the system into an active subsystem, which is the focus of the computational efforts and an environment that is treated at a lower computational level. The system partitioning is mostly based on localized molecular orbitals. When reaction paths or energy differences are to be calculated, it is crucial to keep the orbital space consistent for all structures. Inconsistencies in the orbital space can lead to unpredictable errors in the potential energy surface. While successful strategies to ensure this consistency have been established for organic and even metal-organic systems, these methods often fail for metal clusters or nanoparticles with a high density of near-degenerate and delocalized molecular orbitals. However, such systems are highly relevant for catalysts. Accurate yet feasible quantum-mechanical *ab initio* calculations are therefore highly desired. In this work, we present an approach based on the SPADE algorithm that allows us to ensure an automated and consistent partitioning even for systems with delocalized and near-degenerate molecular orbitals and demonstrate the validity of this method for the binding energies of small molecules on transition-metal clusters.

## 1. INTRODUCTION

In recent years, as worries about climate change and environmental pollution have grown, heterogeneous catalysis has gained increasing attention, since these catalytic transformations are among the most energy consuming industrial processes.[1,2] Improvements in these catalysts' selectivity and efficiency are thus extremely desirable not only from a commercial point of view but also for addressing environmental challenges such as the reduction of greenhouse gas emissions.[3-6] Metal nanoparticles are excellent candidates as catalysts, since they provide a large surface area with a high number of possible active sites, enabling efficient and effective catalytic reactions.[7-10] Since their catalytic abilities rely on various intrinsic and environmental conditions, it is essential to derive a detailed understanding of their electronic properties and operating mechanisms.[11-15]

In this regard, computational models became a key component. The accurate description of complex electronic structures is a significant challenge even for modern computational methods.[12,14,16] Although wave function theories (WFT) such as configuration interaction,[17,18] coupled cluster,[19,20] or perturbative post-Hartree-Fock[21,22] methods allow to achieve highly accurate and systematically improvable results, they are numerically too expensive for large systems.[23,24] In contrast, Kohn–Sham Functional Theory[25] (KS–DFT) provides a computationally more efficient approach at the price of substantial limitations due to the approximation of the exchange functional.[26] The choice of the exchange functional has a significant impact on the accuracy and varies depending on the system. In particular, the description of chemical reactions on transition metal complexes or nanoclusters, which exhibit (near-)degenerate electronic character, are known to be error-prone.[27]

In the last decades, embedding methods have gained increasing attention.[28,29] These methods are designed to precisely describe chemical reactions at low computational cost. This is achieved by separating the system into an active subsystem, which is the focus of computational efforts, and an environment that is treated with a more approximate method. Versatile quantum embedding schemes such as QM/MM,[30] implicit solvents,[31,32] ONIOM models,[33,34] density matrix embedding theory (DMET),[35] Green's function embedding,[36-38] frozen density embedding (FDE) and subsystem DFT[39-46] among many others[47-49] have been developed.

Projection-based embedding theory (PBET) is a subsystem DFT approach, which offers a robust, accurate, and simple framework for DFT-in-DFT and even WF-in-DFT embedding.[46,50] In PBET, the subsystem partitioning is based on conventional orbital localization procedures such as Pipek-Mezey[51] (PM) or intrinsic bond orbitals[52] (IBOs), assigning molecular orbitals (MOs) to atoms, which correspond to the active or environmental subsystem, respectively. The partitioning of the orbital space based on such localization schemes has been termed manual selection (MS).[53] However, during chemical reactions, the nature of the selected MOs might change, and with it, the partitioning of the active and the environmental subsystems. This might lead to unphysical discontinuities in the potential energy surface (PES).[54-56] To overcome this issue, so-called "even-handed" selection schemes have been proposed.[53,57,58] The common idea of these methods is to identify MOs in the active region, which are relevant for describing the chemical reaction, and to form a consensus set of these MOs, which are then used for all embedding calculations along the reaction pathway. This eliminates discontinuities in the PES and provides an accurate evaluation method for energies. Automated approaches, such as the direct orbital selection (DOS) scheme invented by Bensberg and Neugebauer[53,58] have been developed for partitioning the molecular system most efficiently and accurately. We note however that these orbital selection approaches are all based on conventional localization schemes, which tend to strongly localize the MOs such that contributions from only one or two atoms are dominant.[55]

As an alternative, another elegant active orbital space selection scheme has recently been published by Claudino and Mayhall, which has been termed subsystem projected atomic orbital decomposition (SPADE).[55] In this approach, again, a set of atoms is manually selected and allocated to the active subsystem. Based on this selection, the MOs are projected onto orthogonalized atomic orbitals of the active subsystem. Subsequently, a singular value decomposition (SVD) is applied to the subsystem coefficient matrix of the active system. The transformation matrices derived from the SVD allow to transform the canonical MOs to a set of localized SPADE orbitals. The largest gap between consecutive singular values defines the most suitable system partitioning. This approach is, in contrast to the conventional localization schemes, independent of any externally set parameters or thresholds. In contrast to the conventional orbital localization procedures (such as IBO or PM), which produce MOs with almost atomic-orbital character, the SPADE algorithm localizes more broadly obeying the subsystem boundaries.[55] Further, studies have shown that this approach is more robust towards the variation of a reaction coordinate than the conventional localization schemes.[55] However, despite the robustness increased, discontinuities cannot generally be excluded.[59]

Although significant achievements have been made in the development and application of different selection schemes in PBET over the last decade, most of the reported studies focus either on organic systems or on small transition metal complexes.[50,57,60-65] Systematic studies dealing with energetically (near-)degenerated many-electron systems such as metal nanoclusters are so far limited. In particular, the application of the traditional localization schemes on metal nanoclusters is questionable since the MOs of such systems are strongly delocalized. Although there is some evidence, that the regionally localizing SPADE algorithm might be most suitable for partitioning non-localized MOs, a detailed analysis is missing. Therefore, a comprehensive picture of the reliability of different active space selection schemes on metal nanoclusters is required in order to expand the powerful PBET method to this class of systems.

Here, we report a systematic study analyzing the suitability of different active orbital selection approaches for the calculation of PES of small molecules interacting with metal nanoclusters. We show that, due to the delocalized nature of the MOs, orbital selection schemes, which are based on traditional localization schemes, result in enormous energetic discontinuities along the reaction pathway. The SPADE algorithm is by contrast more robust but still suffers from discontinuities in challenging cases. However, we show that this issue can be solved, by utilizing an MO selection based on the singular value hierarchy. Our study shows that the SPADE algorithm consistently prioritizes the most relevant MOs even for (near-)degenerate orbitals, which allows us to trace the evolution of MOs along the reaction pathway. We exploit this finding to select a consistent set of MOs throughout the entire trajectory, converting the SPADE algorithm into an even-handed approach, which successfully eliminates all discontinuities. Although this approach solves the discontinuity problem, the question of the accuracy of this selection approach is still open. Our study demonstrates that simply increasing the number of actively treated molecular orbitals in the active system does not necessarily lead to a more accurate description of the dissociation curve. This behavior prevents systematical improvements and error estimations, which are highly required to increase the predictive power of these QM/QM embedding methods. To overcome this issue, we developed an **A**utomated and **C**onsistent **E**ven-handed partitioning approach for systems with (near-)degenerate MOs based on the SPADE algorithm and termed **ACE-of-SPADE**. We demonstrate the validity of this method for small molecule binding energies on transition metal clusters. The ACE-of-SPADE approach adequately balances the size of the active system and the accuracy.

Our study effectively addresses the challenges regarding system partitioning that come with the application of the PBET method on metal nanoclusters. The developed ACE-of-SPADE algorithm allows automated and efficient selection of active orbitals, which leads to reliable and systematically improvable results. This study shows how PBET can successfully be applied on strongly correlated and (near-) degenerate systems and makes therefore a significant contribution to the computational research of heterogeneous catalysis.

## 2. COMPUTATIONAL DETAILS

For this study, an $Au_{13}$ nanocluster with a bonded carbon monoxide molecule ($CO@Au_{13}$) has been studied as a prototypical test system. Before running the embedding calculations, a geometry optimization has been conducted using Q-Chem 5.4[66] with the PBE0 functional[67] and the def2-SVP basis set.[68] The embedding calculations are performed with an adapted version of the quantum chemistry package Serenity.[69] All embedding calculations are conducted as a Huzinaga-type top-down PBET in frozen-density embedding mode (FDE),[70] using unless otherwise stated, PBE0 for the active subsystem and LDA[25,71] for the environment. The level shift parameter is set to $1.0\ E_h$. For the embedding calculations, the def2-SVP basis set is applied as well. For the manual orbital selection schemes IBO localization is used. The Mulliken net population threshold is set to $\tau_{\text{Mulliken}} = 0.4$ to separate active from environmental MOs. When running the DOS scheme, the automated orbital alignment strategy of Bensberg and Neugebauer is applied.[53] For comparing the MOs, the selection thresholds $\tau_{\text{kin}}$ and $\tau_{\text{loc}}$ must be defined. In this work, they are set to be equal ($\tau_{\text{DOS}} = \tau_{\text{loc}} = \tau_{\text{kin}}$). To modulate the size of the active space, these parameters are varied in a range between 0.05 and 1. The algorithm used to calculate partial charges is based on the shellwise accumulation of intrinsic atomic orbitals (IAOs).[52,53]

The structures along the reaction pathway are created based on the PBE0-optimized CO@Au$_{13}$ complex. The vector pointing from the closest Au-atom to the C-atom defines the direction in which the CO substrate is iteratively moved without modification of the CO bond length (see Figure S6). Starting from a distance of 1 Å between substrate and metal nanocluster the substrate is moved along the predefined vector for 40 steps with a step size of 0.1 Å. Scripts and modifications to Serenity 1.4 required to run our proposed active-space selection algorithm are provided in the SI.

## 3. RESULTS AND DISCUSSION

### 3.1 Assessing the Quality of Active Space Selection Schemes for Metal Nanoclusters

First, we evaluate the quality of different active space selection schemes for the application on metal nanoclusters. Therefore, we compare the dissociation curves of CO@Au$_{13}$ calculated using PBET with different active space selection schemes with the dissociation curve obtained from a pure PBE0 calculation (see Figure 1). An intuitive choice for the system partitioning is to assign the substrate and the closest Au atom to the active subsystem, while all the other Au atoms constitute the environment. The MOs of the active subsystem can then either be obtained from the MS scheme or from the SPADE algorithm. Figure 1a) shows the dissociation curves, which result from such a partition. The MS scheme leads to discontinuities at larger Au-C distances ($r >$ 3.3 Å), due to fluctuations of the active orbital space size between 32 and 33 active MOs. In contrast, the SPADE algorithm is more robust regarding fluctuations of the actively treated MOs, leading to a smooth and continuous energy curve along the entire reaction pathway. However, the binding energy is overestimated by around 0.15 eV with respect to the pure PBE0 reference calculation for both selection schemes. In the next panel (Figure 1b), the number of the actively treated atoms is manually increased, and the next layer of the metal cluster is included in a CO@Au$_5$ active subsystem. For this system partition, both, the MS and the SPADE algorithm show discontinuities. The MS approach fluctuates between 108 to 111 actively treated orbitals leading to pronounced energy fluctuations along the reaction pathway. The origin of these strong fluctuations is the large number of delocalized MOs of the metal nanocluster. Figure S7 shows exemplary MOs after applying the IBO localization procedure. Despite the application of traditional localization schemes, the MOs are still delocalized over the entire system. This, combined with the large number of (near-) degenerate orbitals leads to multiple equivalent localized MOs. Even for slight structural changes, localization might lead to different sets of localized MOs. The unique identification of a consistent set of MOs along the reaction pathway is therefore not possible. The dissociation curve calculated with the SPADE algorithm is more robust. Nevertheless, a large energetic shift occurs at 3.7 Å. At this point, the number of active MOs drops from 110 to 108. Figure 2 illustrates the evolution of the singular value distributions along the reaction pathway (see Figure 2b). In line with the discontinuities in the energy curve, the largest gap between adjacent MOs – which determines the MO selection – changes at 3.7 Å and causes the change of the active space size. Although the position of the largest gap changes, it should be highlighted that this approach still allows tracking the evolution of the MOs. The SPADE algorithm thus provides a consistent identification framework, which is required for the development of an even-handed scheme for many-body systems with many (near-)degenerate orbitals.

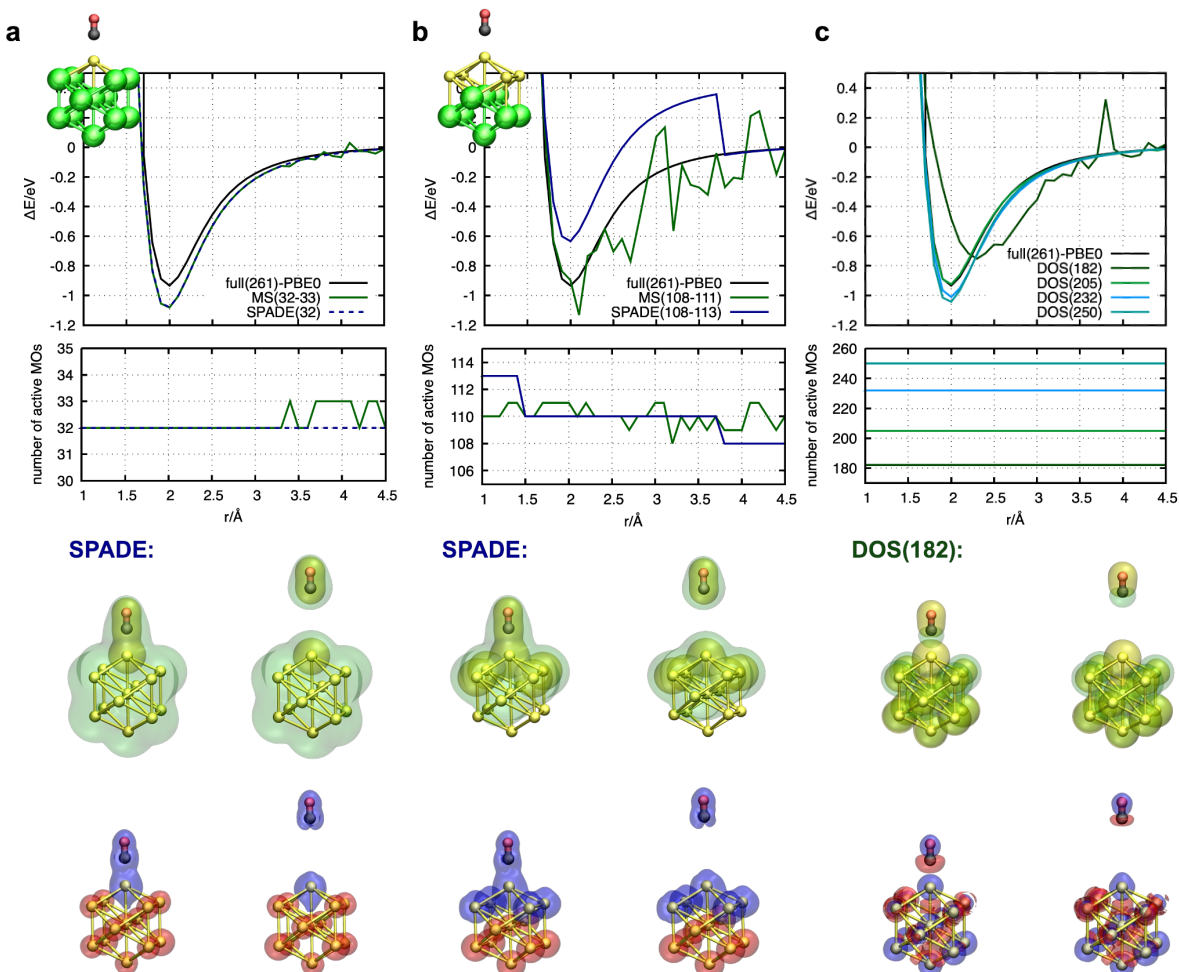

**Figure 1.** Impact of different active space selection schemes on the dissociation curve of CO@Au$_{13}$. a/b) Dissociation curve calculated by embedding a CO@Au/CO@Au$_5$ cluster as active subsystem into the Au$_{12}$/Au$_8$ environment, respectively. c) Dissociation curve using the DOS approach considering different sizes of actively treated molecular orbitals. The density plots below show the density of the active subsystem (yellow) and the environmental subsystem (green) at the isovalue of 0.025 a.u. for two different configurations along the reaction pathway. The left image in each subfigure corresponds to the configuration at $r = 2.0$ Å and the right image at $r = 3.9$ Å. In the bottom plots, the charge differences are depicted, comparing the density distribution of the pure PBE0 calculation with the density distribution after embedding: $(\rho_{\text{diff}} = \rho_{\text{pure-PBE0}} - \rho_{\text{PBE0-in-LDA}})$. The blue areas highlight positive charge accumulation, while the red areas highlight negative charge depletion zones. The isovalues are 0.1 in a) and b) and 0.001 in (c).

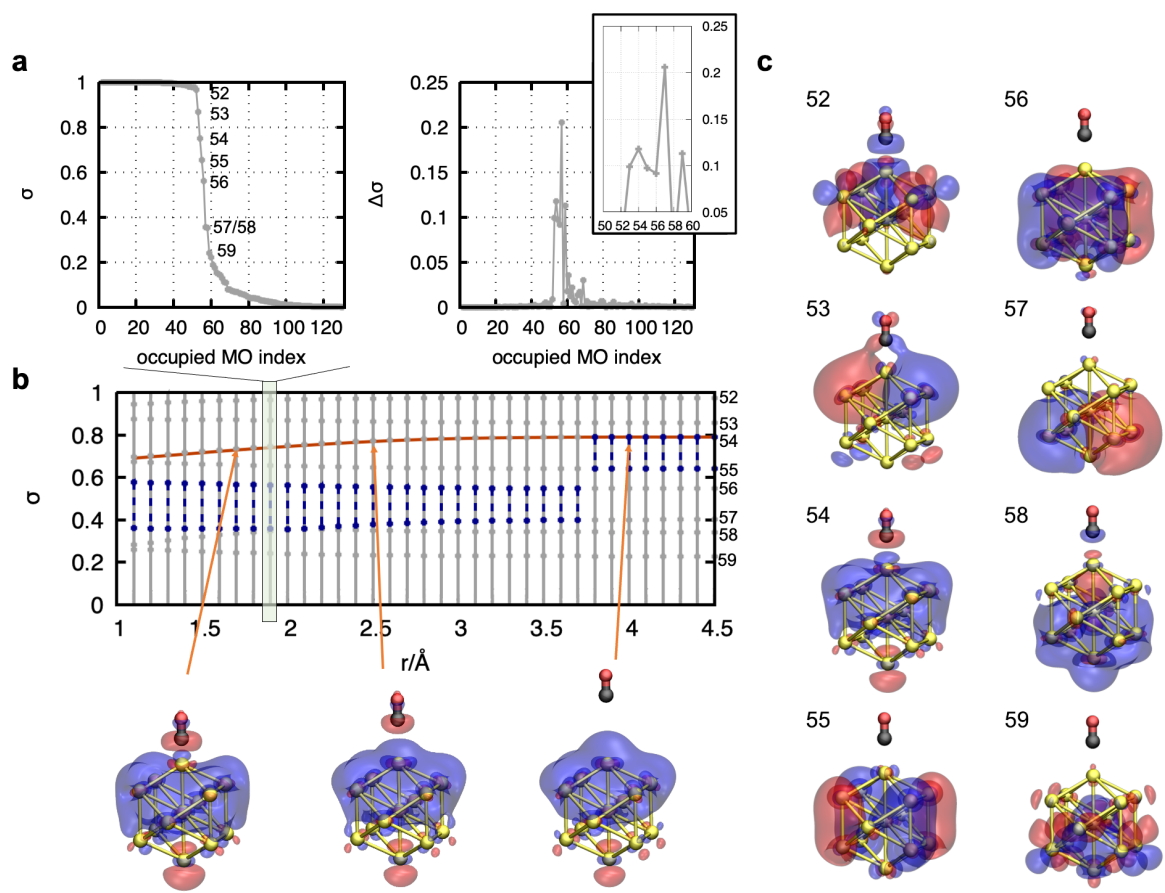

**Figure 2.** a) Plots of the singular value $\sigma_i$ distribution and the difference of consecutive MOs with respect to the MO index ($\Delta\sigma_i = \sigma_i - \sigma_{i-1}$) for the CO@Au5-in-Au8 nanocluster. The inset in the right plot shows a zoom-in into the multi-peak structure of the $\Delta\sigma_i$ curve. b) Evolution of the singular values along the reaction pathway. At each point $r$ of the reaction coordinate the singular values of MO with index 52-59 are shown by the gray dots. The gray lines connecting the singular values are printed as a guide to the eye, highlighting the affiliation of the singular values to the respective reaction coordinate. The blue-colored line fragments mark the largest gap between adjacent MOs, used in automated SPADE for system partitioning. The evolution of the MO with index 54 (highlighted by the orange curve) is shown by the three exemplary plots below. The Au atoms are colored yellow, the C atom black, and the O atom in red. The alpha MOs are shown by the blue) and beta MOs by the red isosurfaces with isovalues of 0.015 and -0.015, respectively. c) Illustration of the MOs with index 52 to 59 at r=2.0 Å. The same color coding as in b) is used.

This aspect will be discussed in more detail in section 3.2. Apart from this, Figure 1c) illustrates the performance of the DOS approach, used to enforce an even-handed localization protocol based on conventional localization methods. In contrast to the previously discussed active space selection approaches, a predetermination of active atoms is not required here. In this approach active MOs are selected automatically based on the changes in their composition over the reaction trajectory. After the identification of relevant orbitals, this consensus set is used for all embedding calculations. Although such an even-handed approach should promise a smooth and continuous energy potential landscape significant problems arise when applied to the metal nanocluster. Using recommended thresholds ($\tau_{DOS} = 0.05$)[53] for the comparison algorithm leads to active spaces in which almost all MOs are considered. This effect arises from the previously discussed observation that the localization schemes might converge to different sets of MOs for slightly modified structures. These inconsistencies are accidentally associated with MO changes along the reaction coordinate. Consequently, all of these MOs are considered as active. The resulting curves are smooth, but the advantage of increased computational efficiency due to small active MOs spaces is lost. In this regard, increasing the threshold allows to reduce the size of the active system. Therefore, unconventional parameters up to $\tau_{DOS} = 1$ are applied. This reduces the active orbital size only to 182 MOs, which is still more than half of all MOs. Further, with such large thresholds, the even-handedness of the selection procedure fails, which is manifested in several discontinuities along the reaction curve. This has been observed previously for large thresholds in the DOS procedure.[53]

The analysis has shown that active space selection schemes based on localization algorithms such as IBO, which produce strongly localized orbitals, introduce large errors when analyzing molecular properties of metal nanoclusters along a reaction pathway. The SPADE algorithm is less error prone since it localizes more broadly respecting the selected active subsystem, and therefore keeping the delocalized character of the molecular orbitals to some degree. These properties can also be observed by analyzing the charge distributions of the active and environmental systems. Figure 1a-c) (bottom panel) shows the density distribution of the active subsystem (yellow) and the environmental subsystem (green), obtained from the SPADE approach (a,b) and the DOS approach (c). Interestingly the density distributions of the molecular environment obtained from the SPADE approach cover the entire molecular systems. The molecular density of the environment around the active region is caused by the delocalized nature of the molecular orbitals. Although the orbitals which contribute most to the active system are identified and considered as active, the accumulation of the minor contributions of the remaining delocalized orbitals assigned to the environment results in this density distribution. However, as shown in the charge difference plot (bottom row in Figure 1), which compares the total density of both subsystems with the target density obtained from the full PBE0 reference calculation, the accumulated density of both systems around the active region is overestimated, while the electron density in the environment is depleted.

### 3.2 Active Space Convergence

The SPADE methodology enables a largely robust and accurate selection of the active MO space. Further, the SPADE algorithm retains some degree of delocalization for the orbitals of the active subsystem and introduces a ranking based on the singular values. This allows to analyze the evolution of MOs along the reaction pathway and provides a possibility for uniquely identifying the relevant MOs over the course of a trajectory. This can be exploited to derive an even-handed approach based on the SPADE algorithm.[59] However, embedding calculations promise to balance high accuracy and computational efficiency. Therefore, it is crucial to investigate the convergence behavior between the selected active space size and the resulting errors. To study this, the dissociation curve is calculated by applying the SPADE algorithm on two different active clusters (CO@Au-in-Au$_{12}$ and CO@Au$_5$-in-Au$_8$). For both cluster configurations, different sizes of active MO sets are considered. Along each curve, the number of the MOs is kept fixed (see Figure 2a)). The active orbitals are selected with respect to their index and ranked according to their singular values. Contrary to our expectations, increasing the active space size does not necessarily increase the accuracy of the embedding calculation. In the case of the CO@Au-in-Au$_{12}$ embedding calculation discontinuities occur for large active spaces, indicating that the selection approach is no longer even-handed. The size of the hierarchically ordered MOs depends on the number of active atoms. In the case of CO@Au-in-Au$_{12}$, only 118 MOs are considered during the singular value decomposition. Selecting MOs beyond this limit leads to discontinuities, since they are not included in the localization scheme. Further, even calculations based on active space sizes lower than 118 might be error prone. Although MOs are prioritized by the singular value decomposition, many MOs have singular values close to 0 (see Figure S8)). For such low numbers, the ordering is inconsistent between different structures, which results in fluctuations of the energies along the dissociation curve (see the energy curve based on 92 active MOs in Fig. 3a)). The best result for the CO@Au-in-Au$_{12}$ configuration is achieved by selecting the active space with respect to the largest gap between consecutive MOs, leading to a constant active orbital space with 32 active MOs. The situation for the CO@Au$_5$-in-Au$_8$ (Figure 3b) configuration is similar. Increasing the number of active MOs may reduce the accuracy. The best result is obtained for 108 active MOs, which corresponds to the average number of active MOs selected by SPADE along the trajectory.

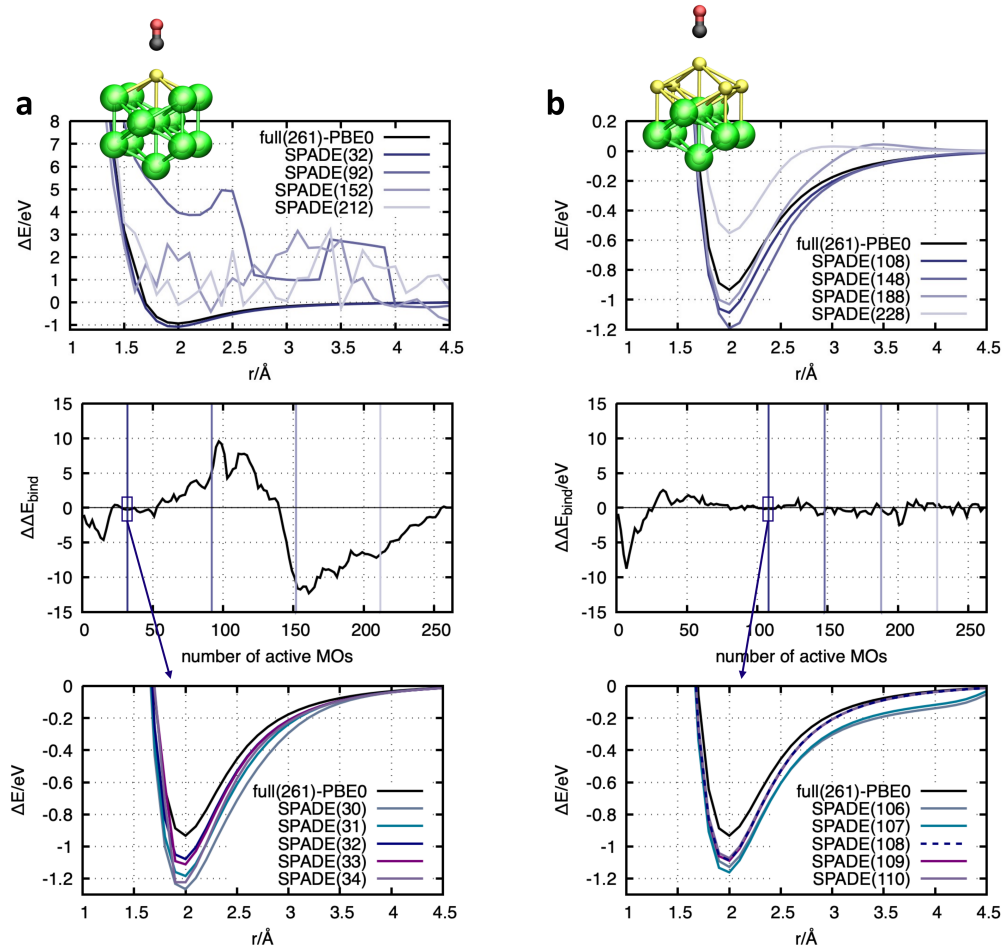

**Figure 3.** Impact of the active space selection on the binding energy $\Delta E$ between an $Au_{13}$ nanocluster and a CO molecule. The top image shows the binding energy $\Delta E$ with respect to the reaction coordinate $r$, which measures the distance between the C atom and closest Au atom of the nanocluster. The insets show the CO@$Au_{13}$ nanocluster. The green-colored Au atoms are considered as molecular environment. The yellow Au atoms and the black C and red O atoms are treated as active. Orbitals of the environmental system are optimized on LDA and orbitals of the active system on the PBE0 level. The panel in the middle shows the error in the binding energy ($\Delta\Delta E_{bind} = \Delta E^{min} - \Delta E^0$) calculated based on the energies of configuration at $r$=2.0 Å ($\Delta E^{min}$) and $r$=3.9 Å ($\Delta E^0$). Only configurations with a multiplicity of 1 or 2 are considered. The vertical lines highlight the active space used for calculating the binding energy curve above. The bottom panels show dissociation curves calculated based MOs for a small window around the original SPADE partition.

This study shows that the active space sizes can be categorized into three regimes: In the first regime, an active space selection is attempted for orbitals with $0 < \sigma_i < 1$. This is the region where the automated SPADE algorithm partitions the systems. Keeping the number of MOs in this regime constant leads to an even-handed orbital selection scheme. In the second regime, the active space partitioning is applied on MOs that have similar singular values very close to 0. Since small (numerical) fluctuations can change the MO ordering, it cannot be ensured that a consensus set of the same MOs is used along the trajectory. The last regime for active MO sizes is an attempt to select orbitals that were not subjected to the singular value decomposition. The MOs in this regime are not localized. Consequently, the uniformity of MOs according to the singular value hierarchy cannot be established.

To get further insights into the dependence between the accuracy of the binding energy and active space size, the errors of the binding energy with respect to full PBE0 calculation have been calculated (see Figure 3, central plot). The plots for both systems CO@Au-in-Au$_{12}$ and CO@Au$_5$-in-Au$_8$ show a non-monotonic and non-converging behavior. The lowest errors are obtained from a small range corresponding to the first regime (see highlighted rectangles in Figure 3, central plot). For CO@Au-in-Au$_{12}$, the best result in this transition area ($0 < \sigma_i < 1$) is achieved by following the largest gap selection of the original SPADE (see Figure 3a), bottom). Note, even small deviations in the active space size ($\pm 1$ active MO) affect the accuracy significantly (see Figure 3a), bottom). For the CO@Au$_5$-in-Au$_8$ a selection based on the largest gap is not straightforward, since the gap position changes along the trajectory. Due to the active atom selection partitioning the covalently bound metal nanocluster, many MOs have singular values in the transition region. Here, the best partitioning is obtained by the average position of the largest gaps along the trajectory. In this way, all the most favorable partitions of all structures are taken into account (see the dissociation curve in Figure 3b), bottom panel, calculated based on 108 active MOs). Analogous to the CO@Au-in-Au$_{12}$ configuration, small fluctuations in the size of active orbital spaces ($\pm 1$ active MO) lead to an adaption of the energy profile (see Figure 3b) bottom). However, the deviations in the errors are smaller compared to CO@Au-in-Au$_{12}$, due to the larger active system defined by the predefined active atoms. The individual MOs contribute less to the CO interaction of the active system, due to the extended cluster size. Additionally, it should be mentioned that with increasing number of active atoms, the errors of the different energy contribution terms decrease. This observation makes the approach more robust towards errors due to slight changes in the number of actively treated MOs. This will be discussed in more detail in section 3.4.

This analysis shows that despite the prioritization of MOs according to the SPADE algorithm, no convergence with respect to the active system size is achieved. In contrast, we could show that the most reliable results, concerning consistency and accuracy, can be obtained from a small window around the singular value transition region. In this regime, the space sizes are comparably small. This is promising since high accuracy for small active system sizes is one key requirement for successful embedding strategies.

### 3.3 ACE-of-SPADE – Automated, Consistent, and Even-Handed Active Space Selection Scheme

The SPADE algorithm bridges the gap between accuracy and efficiency as shown above and provides moreover a suitable framework to establish an even-handed approach. However, due to missing convergence with respect to the active space size, the selection of appropriate consensus sets of MOs along the trajectory is still challenging. As shown above, the most reliable results are obtained by focusing on MO selections close to the singular value transition region. When dealing with metal clusters, where a lot of MOs with similar singular values in the transition area exist, the unique partitioning based on the largest singular value gap becomes error prone. On the one hand, the position of the largest gap along the reaction trajectory can change, which leads to unphysical discontinuities in the PES. On the other hand, even small deviations from the transition region reduce the accuracy.

To solve these problems we developed an **A**utomated, **C**onsistent, and **E**ven-handed active space selection scheme based on the SPADE algorithm, called **ACE-of-SPADE**, for the successful application of PBET on metal nanoclusters and similarly challenging systems. This algorithm addresses the above-mentioned issues by performing iteratively three different steps (see Figure 4): In the first step, the SPADE algorithm is applied to each individual structure along the reaction coordinate to get the information about the prioritization of MOs based on the singular values. In the second step, the first derivative of the singular value distribution curve is fitted to the derivative of the square root of the Fermi function $\Delta\sigma'^{\uparrow/\downarrow}_{RC}(i) = \frac{\beta e^{\beta(i-MO^{\uparrow/\downarrow}_{max,RC})}}{\left(1+e^{\beta(i-MO^{\uparrow/\downarrow}_{max,RC})}\right)^{1.5}}$, for each structure along the reaction coordinate separately. The variable $i$ labels the MO index and $MO_{\max,RC}$ describes the peak position of the fit function obtained from the singular value distribution at the reaction coordinate $RC$. Note, this analysis is performed independently for the alpha and beta MOs, indicated by ↑/↓, respectively.

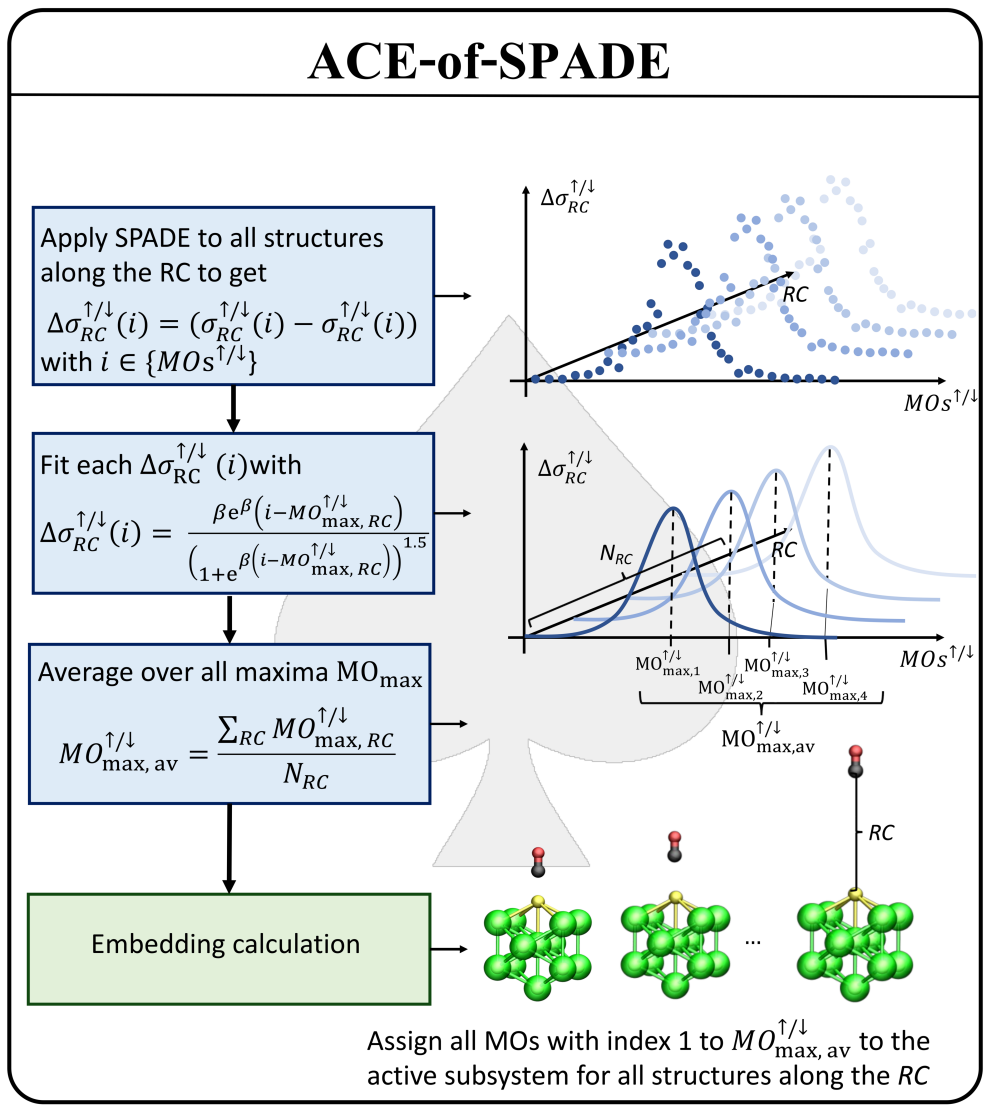

**Figure 4.** Flowchart of the ACE-of-SPADE active orbital space selection scheme for calculating potential energy surfaces of chemical reactions. The images beside the flowchart illustrate the respective steps.

The introduction of the fit function leads to more stability regarding the system partitioning, in particular for systems where many singular values are in the transition area between $0 < \sigma_i < 1$. In these cases, changes of the largest gap position along a reaction coordinate are more likely to occur. Due to the fitting procedure, the peak position depends not only on the position of individual singular values but on the distribution of all singular values close to the transition area. Thus, the peak of the fit function indicates the most suitable system partition. All MOs with an index lower than that determined by the peak are assigned to be the active subsystem, while all the other MOs belong to the environment. The motivation for exploiting this functional form of the fit function comes from the analogy of interpreting the squared singular values similar to occupation numbers.[55] Although this approach is more stable for metallic systems than the automated active space selection according to the largest gap of consecutive singular values, it can still not generally be excluded that the peak position changes distinctively along the trajectory. To compensate for this case, in a third step, the average value of all peak positions is calculated. This final average value indicates where the system is partitioned, with respect to the requirements of the individual configuration along the whole trajectory. The last step ensures the uniformity of taking the same set of MOs at each point of the reaction coordinate into account.

This method ensures even-handedness as long as the hierarchy of the MOs defined by their singular values does not significantly change along the reaction coordinate. Although such a phenomenon has not been observed so far, it is still conceivable. This is problematic because a changing hierarchy can destroy the selection of a consensus set of the same MOs with respect to their ascending MO indexes. However, the (ACE-of-) SPADE provides a framework to address this issue. As shown in Figure 2, the evolution of the MOs can be traced and uniquely identified. If singular values of MOs close to the transition area change dramatically along the RC, such that the hierarchy changes, the corresponding MOs must be taken into account regardless of their final MO index. In this way, even-handedness is re-established. However, a detailed accuracy analysis must be carried out if such a phenomenon will be observed in future studies.

### 3.4 Application and Verification: Convergence and Error Cancellation

In order to demonstrate the reliability and validity of the ACE-of-SPADE active orbital space selection, the dissociation curve of CO@Au$_{13}$ has been investigated by considering six different active atom selections (see Figure 5a)). These clusters are calculated based on the two different functionals for the environmental system LDA (see Figure 5b)) and PBE[72] (see Figure 5c)). The application of the ACE-of-SPADE selection scheme leads to smooth curves without any discontinuities for all configurations. The analysis of the binding energy with respect to the reference curve on full PBE0 level shows only errors lower than 0.15 eV. Moreover, due to the consistent selection scheme, a convergence of the accuracy with respect to the size of the active orbital space is generally achieved. Rather than due to fortuitous error cancellation, the convergence is caused by decreasing errors of the individual energetic contributions, i.e. from the active system, the environment and the interaction terms.

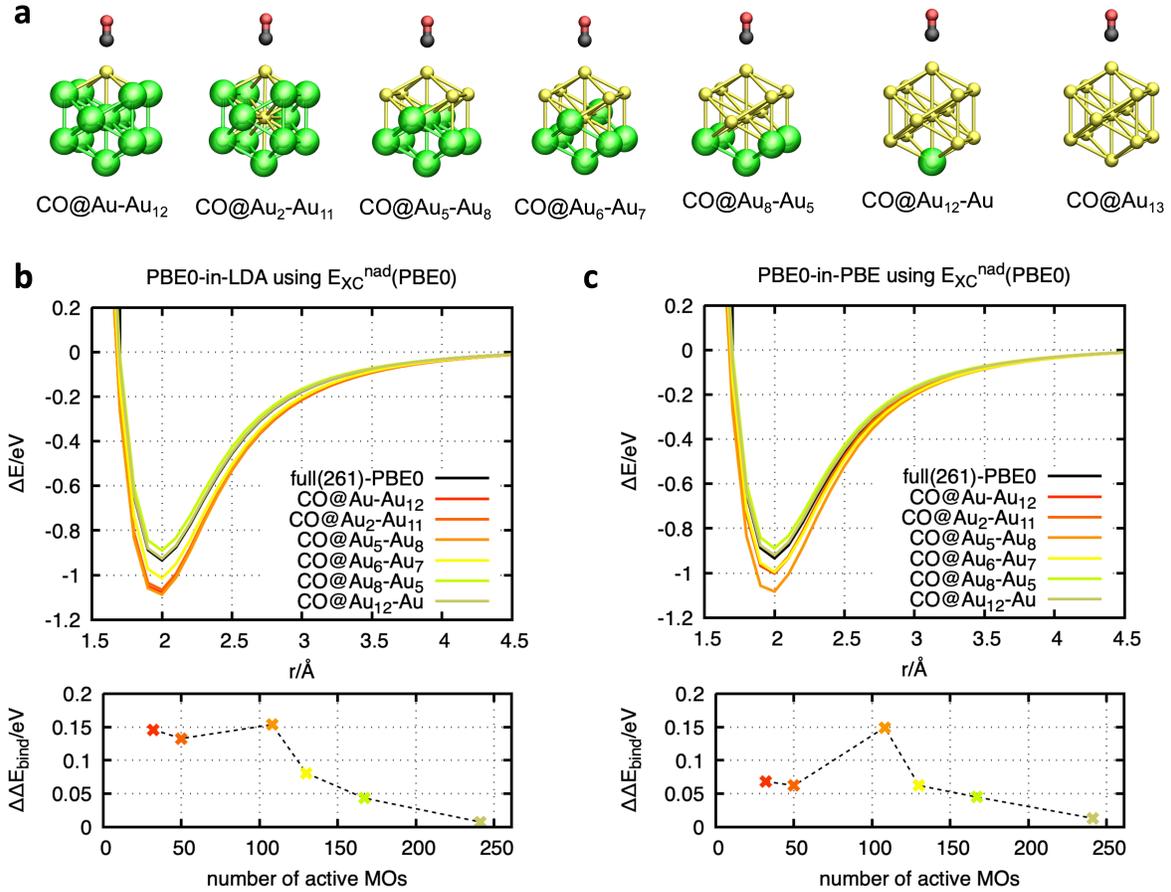

**Figure 5.** Analyzing the influence of the active cluster size on the binding energy. a) Illustration of the different nanoclusters. Green-colored Au atoms are considered as molecular environment and the yellow Au atoms, black C atom, and red O atom describe the active core cluster. b/c) Binding energy along the RC for the different clusters shown in **a**). The top graphs show the dissociation curves, while the bottom plots show the error of the binding energy of the embedding method with respect to full PBE0 reference ($\Delta\Delta E_{\text{bind}} = \Delta E_{\text{bind}}^{\text{ACE-of-SPADE-PBET}} - \Delta E_{\text{bind}}^{\text{PBE0}}$). The plots in **b**) are based on PBE0-in-LDA embedding, using PBE0 for the non-additive exchange functional. The plots in **c**) are based on PBE0-in-PBE embedding using PBE0 for the non-additive exchange functional.

The error of each contribution is calculated according to $\Delta\Delta E_X(r) = \Delta E_X^{\text{ACE-of-SPADE-PBET}}(r) - \Delta E_X^{\text{PBE0}}(r)$, while $X$ labels the respective contribution: $X \in \{\text{active}, \text{environment}, \text{interaction}\}$. The contributions for $\Delta E_X^{\text{PBE0}}(r)$ are obtained from PBE0-in-PBE0 embedding calculations based on the same system partition as the ACE-of-SPADE embedding calculation. Figure 6 shows the errors of the different contributions for the three different active system sizes CO@Au-Au$_{12}$, CO@Au$_7$-Au$_5$ and CO@Au$_{12}$-Au. The top image of each subfigure shows the error contributions and their total accumulation for considering the environment with the LDA functional and the bottom image shows the errors for a calculation where the environment is treated with the PBE functional. This analysis shows that the surprisingly high accuracy, achieved with small cluster sizes, results from error calculations of large negative errors $\Delta\Delta E_{\text{active}}(r)$, with smaller positive errors of both the environment $\Delta\Delta E_{\text{environment}}(r)$ and the interaction contributions $\Delta\Delta E_{\text{interactions}}(r)$ (see Figure 6a,b)). The larger the active space, the smaller the individual error contributions (see Figure 6c). Thus, increasing accuracy for increasing active systems is not coincidental but a result of decreasing individual errors as should be the case.

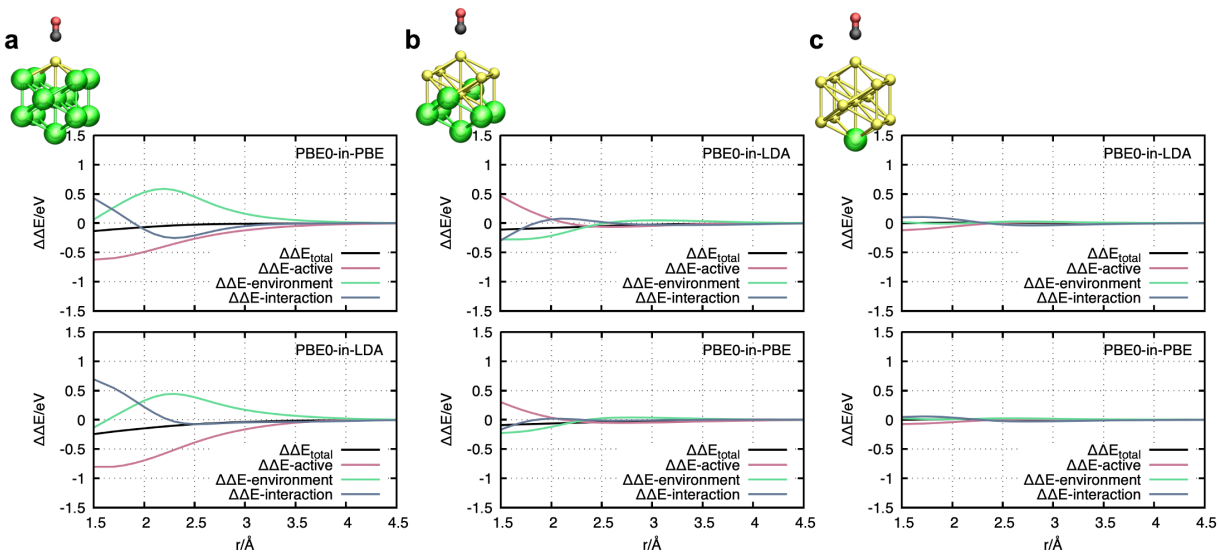

**Figure 6.** Individual error contributions for different systems with different active atom configurations. The black curves represent the total error given by the accumulation of individual contributions. The top diagrams show the calculated errors for treating the environment with the LDA functional, while the diagrams show the environment with the PBE functional. a) shows the errors for the cluster partitioning CO@Au-Au$_{12}$, b) for CO@Au$_7$-Au$_6$ and c) for CO@Au-Au$_{12}$.

However, despite the observed convergence, the CO@Au$_5$-Au$_8$ cluster configuration constitutes an outlier in the dissociation curves shown in Figure 5. For this active cluster size, the individual error contributions are sufficiently large, such that mutual error compensation is required for a (fortuitous) accurate result (see Figure S9). Apparently, due to a bad definition of active atoms, the error cancellation is poor-balanced, which leads to total errors of around 0.15 eV.

The observed convergence behavior as well as the generally low errors and the smooth and continuous curves highlight the validity of the ACE-of-SPADE approach for the application of PBET on systems as challenging as metal nanoclusters. However, errors for small systems that originate from bad active atom definitions cannot be avoided completely, as discussed for the CO@Au$_5$-Au$_8$ partitioning.

### 3.5 Influence of the Non-Additive Exchange Functional in PBET

Analyzing results from calculations with different choices for the non-additive exchange functional indicate that further improvements might be possible (see Figure 7). When using the lower-ranked functional to describe the electron-electron exchange interactions between the active system and the environment, the accuracy for the smallest two active cluster configurations (CO@Au-Au$_{12}$ and CO@Au$_2$-Au$_{11}$) increases dramatically. The errors are reduced to values lower than 0.05 eV (see Figure 7a)) for using the LDA functional, while the PBE functional yields even errors tending towards zero for the 0.0 eV CO@Au-Au$_{12}$ (see Figure 7b)). Such reduced errors, particularly for the smallest clusters, are highly promising in the context of quantum embedding strategies because they indicate a high potential for dramatically reducing the numerical costs without loss of accuracy.

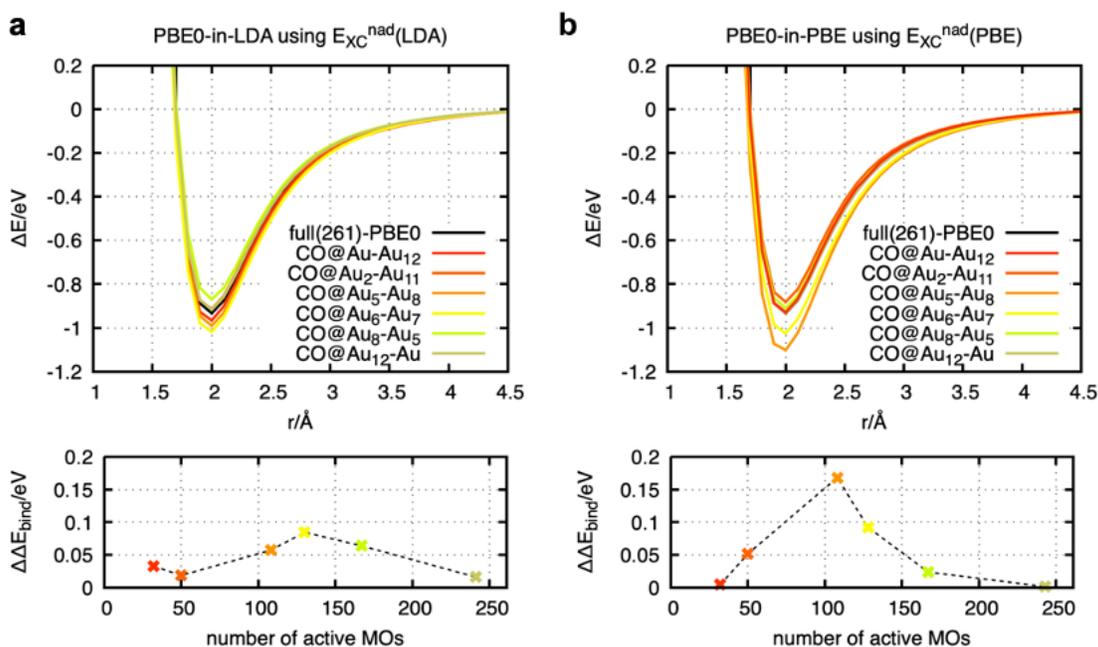

**Figure 7.** The top panels show the dissociation curves for different cluster configurations and the bottom panels show the error of the binding energy of the embedding method with respect to full PBE0 reference calculation ($\Delta\Delta E_{\text{bind}} = \Delta E_{\text{bind}}^{\text{ACE−of−SPADE−PBET}} - \Delta E_{\text{bind}}^{\text{PBE0}}$). The plots in **a**) are based on PBE-in-LDA embedding, using LDA for the non-additive exchange functional. The plots in **b**) are based on PBE0-in-PBE embedding using PBE for the non-additive exchange functional.

The analysis of the individual error contributions indicates that the reduced errors come from well-balanced error cancellation. The individual contributions are larger compared to the conventional PBET using the higher level of theory of the non-additive exchange functional (see Figure S 10). Although the error cancellation is well-balanced for the smallest systems, the errors for larger clusters such as CO@$Au_5$-$Au_8$ and CO@$Au_6$-$Au_7$ are slightly increased (see Figure 7). These systems are still in a regime, where a low total error is due to error cancellation of the individual energy contributions. Since the error compensation is poor for these two clusters, the derivation from reference curves increases and no convergence with respect to the cluster size is observed.

In summary, the dramatically reduced errors for small active subsystems in a metallic environment with the LDA functional for the non-additive exchange contribution is a very promising observation. Further studies for a large selection of metallic clusters are currently performed in our group.

## 4. CONCLUSION

In summary, we presented a systematic study on the application of projection-based embedding theory on metal nanoclusters. Our work highlights and analyzes inconsistencies and inaccuracies, which arise from conventional active space selection schemes. We have shown that active space selection approaches which are based on strict localizations, which tend to converge to molecular orbitals with near-atomic character, cannot adequately partition the molecular system into active and environmental subsystems for a reaction coordinate. The SPADE algorithm is more robust since it preserves the delocalized character of molecular orbitals of metal nanoclusters within the respective subsystem. However, we demonstrated that the selection criterion of the SPADE algorithm becomes problematic for metal nanoclusters due to the high number of (near-)degenerate molecular orbitals and that the automated selection according to the largest gap between consecutive singular values can lead to discontinuities in the energy profile along a reaction coordinate. Nevertheless, the application of the singular value decomposition procedure on the active subspace allows us to reliably rank the molecular orbitals and enables us to follow the evolution of the molecular orbitals along a reaction coordinate. This can be exploited to define a consensus set of molecular orbitals as active orbital space, transforming the automated SPADE into an even-handed approach. The accuracy of the resulting potential energy surfaces depends strongly but non-monotonic on the size of the active orbital space.

To address these issues, we developed an automated, consistent, and even-handed active space selection approach based on the SPADE algorithm (ACE-of-SPADE) that allows the successful calculation of potential energy surfaces of metal nanoclusters by using embedding approaches. The ACE-of-SPADE algorithm considers the distribution of the singular values from each point of the trajectory. In order to avoid inconsistencies from a large number of (near-)degenerate molecular orbitals, the system partitioning is conducted with respect to Fermi-type fit functions. Based on this partitioning, a set of molecular orbitals is selected, which contributes on average most to the active system along the reaction coordinate and ensures both high accuracy and consistency.

We demonstrated the validity of this approach by applying it to a metal nanocluster with different definitions of active atoms. We showed that ACE-of-SPADE provides smooth and continuous energy reaction profiles. Further, our approach establishes converging improvement in accuracy as the active orbital space sizes increase. This observation is confirmed by a detailed analysis of the individual error contributions. With increasing active system sizes, individual errors decrease, which is an indicator of reliable embedding. However, in case of a poor selection of the active region, this convergence can be disturbed particularly for small active systems. There, the individual error contributions are considerably large, but the fortuitous error cancellation is poor. By contrast, we could also show that favorable error cancellation can result in high accuracy at low numerical effort. But in order to exploit advantages from favorable error cancellation further investigations on the impact of the selection of the active region are required. Even though, we stress that the errors for all calculations are relatively small with values lower than 0.15 eV for the binding energy.

Our work introduces the ACE-of-SPADE approach as a reliable active space selection approach for analyzing the potential energy surface of chemical reactions on metal nanoclusters and thus significantly contributes to the computational chemistry toolbox for heterogeneous catalysis.

## ASSOCIATED CONTENT

**Supporting Information**

Additional Information about the modification of the Serenity 1.4 code, detailed explanation about the implementation of ACE-of-SPADE, definition of the probe system, visualization of MOs localized using conventional localization methods, singular value distribution of the CO@Au-Au$_{12}$ cluster, supplementary error compensation plots to Figure 5 and Figure 7.

## AUTHOR INFORMATION

**Corresponding Author**

Christopher J. Stein – Technical University of Munich, TUM School of Natural Sciences, Department of Chemistry, Lichtenbergstr. 4, D-85748 Garching, Germany;

orcid.org/0000-0003-2050-4866

Email: christopher.stein@tum.de

**Author**

Elena Kolodzeiski – Technical University of Munich, TUM School of Natural Sciences, Department of Chemistry, Lichtenbergstr. 4, D-85748 Garching, Germany;

orcid.org/0000-0002-9698-3886

**Notes**

The authors declare no competing financial interest.

**ACKNOWLEDGMENT**

Financial support by the DFG under Germany's Excellence Strategy - EXC 2089/1-390776260 (e-conversion) is gratefully acknowledged. The initial stage of the project was supported by an NRW Return Fellowship to CJS. Computations have been carried out using computational and data resources provided by the Leibniz Supercomputing Centre (www.lrz.de)

**TOC:**

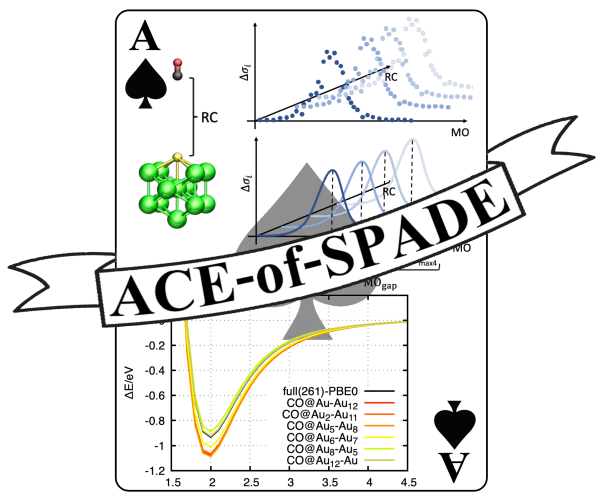

# Supporting Information

# Automated, Consistent, and Even-handed Selection of Active Orbital Spaces for Quantum Embedding

*Elena Kolodzeiski* and *Christopher J. Stein*

Technical University of Munich, TUM School of Natural Sciences, Department of Chemistry, Lichtenbergstr. 4, D-85748 Garching, Germany

**Contents**

1. **Serenity Code Modifications**

In order to manually select where to partition the active and environmental system, modifications on the SPADE algorithm as implemented in the open-source code Serenity 1.4 (https://github.com/qcserenity/serenity/releases/tag/1.4.0) were required. Therefore, the file SPADEAlgorithm.cpp in the directory SERENITY_DIR/src/analysis/orbitalLocalization/SPADEAlgorithm.cpp has been changed. The main modification is given by reading and writing the "LastActiveOrbital_up.csv" and "LastActiveOrbital_down.csv" files. These files contain the integer number of actively treated orbitals for the alpha and beta MOs, respectively. Another small adaption that should be mentioned is made by writing out the singular value distribution. The output files are named "sigma_orb_up.txt" and "sigma_orb_down.txt". The modified code is given below in Figure S1. The modified parts are highlighted in green.

The above-mentioned files are required for conducting the ACE-of-SPADE algorithm. In the following, our implementation of the ACE-of-SPADE approach will be presented. The calculation can be split into three parts. In the first step, for all structures along the reaction pathway, the SPADE algorithm is applied with the help of the Serenity program. In Figure S2 an exemplary serenity input is shown. With this input file, the Serenity program conducts the SPADE analysis for one given configuration and writes out the singular value distribution for the alpha and beta MOs ("sigma_orb_up.txt" and "sigma_orb_down.txt"), due to our modification. In the second step, post-processing takes place, extracting information about the preferred system partitioning from the "sigma_orb_up.txt" and "sigma_orb_down.txt" files. To do so, all obtained singular value distributions are fitted with the first derivative of the square root of the Fermi-type function (see Figure 3). This is carried out with the python script "fit_curve.py" (see Figure S3, tested using python 3.11). This script writes out the peak position of the different structures into individual "lastActiveOrbital_up.csv" and "lastActiveOrbital_down.csv" files. Subsequently, these files are read by the script "calc_av_of_peaks.py" (see Figure S4). This script reads out all the peak positions, calculates the average value and writes it to new "lastActiveOrbital_up.csv" and "lastActiveOrbital_down.csv" files. When stored in the directory, where the serenity calculation will be performed, this file allows the manipulation of the system partitioning with respect to the integer numbers stored in the *.csv files in a subsequent serenity embedding calculation. The latter is the last step. Based on the "lastActiveOrbital_up/down.csv" files, the actual embedding calculation can be performed for all structures along the reaction coordinate. One representative serenity input file, required for this, is shown in Figure S5.

```cpp
#include "analysis/orbitalLocalization/SPADEAlgorithm.h"
#include "basis/BasisFunctionMapper.h"            //Block selection in coefficient matrix.
#include "data/OrbitalController.h"                //Coefficients and orbital update.
#include "data/matrices/CoefficientMatrix.h"       //Coefficient matrix.
#include "data/matrices/MatrixInBasis.h"           //Overlap matrix.
#include "integrals/OneElectronIntegralController.h" //Overlap integrals for symmetric orthogonalization.
#include "io/FormattedOutputStream.h"              //Filtered output streams.
#include "math/linearAlgebra/MatrixFunctions.h"    //Matrix sqrt
#include "misc/SerenityError.h"                    //Error messages
#include "system/SystemController.h"               //System controller defintion.
#include <fstream>
namespace Serenity {
template<Options::SCF_MODES SCFMode>
SPADEAlgorithm<SCFMode>::SPADEAlgorithm(std::shared_ptr<SystemController> supersystem,
  std::shared_ptr<SystemController> activeSystem): _supersystem(supersystem), _activeSystem(activeSystem) {}

template<Options::SCF_MODES SCFMode> SpinPolarizedData<SCFMode, Eigen::VectorXi> SPADEAlgorithm<SCFMode>::run()
{
  auto orbitalController = _supersystem->getActiveOrbitalController<SCFMode>();
  const MatrixInBasis<RESTRICTED>& S = _supersystem->getOneElectronIntegralController()->getOverlapIntegrals();
  const Eigen::MatrixXd sqrtS = mSqrt_Sym(S);
  CoefficientMatrix<SCFMode> coefficients = orbitalController->getCoefficients();
  auto nOcc = _supersystem->getNOccupiedOrbitals<SCFMode>();
  SpinPolarizedData<SCFMode, Eigen::VectorXi> assignment;
  BasisFunctionMapper basisFunctionMapper(_supersystem->getBasisController());
  auto projection = basisFunctionMapper.getSparseProjection(_activeSystem->getBasisController());
  std::ofstream outfile1("sigma_orb_up.txt");
  std::ofstream outfile2("sigma_orb_down.txt");
  int spinstate=0;
  for_spin(coefficients, nOcc, assignment) {
    const Eigen::MatrixXd oldC = coefficients_spin.leftCols(nOcc_spin).eval();
    const Eigen::MatrixXd d = oldC * oldC.transpose();
    const Eigen::MatrixXd orthoC = sqrtS * oldC;
    const Eigen::MatrixXd orthoC_A = *projection * orthoC;
    // Calculation of the full matrix V is necessary for the orbital rotation.
    Eigen::JacobiSVD<Eigen::MatrixXd> svd(orthoC_A, Eigen::ComputeFullU | Eigen::ComputeFullV);
    const Eigen::MatrixXd rightSingularVectors = svd.matrixV();
    const Eigen::VectorXd singularValues = svd.singularValues();
    const Eigen::MatrixXd newC = (oldC * rightSingularVectors).eval();
    const Eigen::MatrixXd d2 = newC * newC.transpose();
    if ((d - d2).array().abs().sum() > 1e-10)
      throw SerenityError("Density matrix changed during SPADE orbital construction.");
    coefficients_spin.leftCols(nOcc_spin) = newC;
    unsigned int lastActiveOrbital = nOcc_spin;
    double largestDifference = 0.0;
    for (unsigned int iOrb = 1; iOrb < singularValues.size(); iOrb++) {
      double singularValueDifference = std::fabs(singularValues[iOrb - 1] - singularValues[iOrb]);
      if (spinstate==0){ outfile1 << iOrb << "   " << singularValues[iOrb] << "   " << singularValueDifference << "\n";}
      if (spinstate==1){ outfile2 << iOrb << "   " << singularValues[iOrb] << "   " << singularValueDifference << "\n";}
      if (singularValueDifference > largestDifference) {
        largestDifference = singularValueDifference;
        lastActiveOrbital = iOrb;
      }
    } // for iOrb
    assignment_spin = Eigen::VectorXi::Constant(nOcc_spin, 1);
    if (spinstate==0){
         std::ifstream ifile2("lastActiveOrbital_up.csv");
         if (ifile2){ ifile2 >> lastActiveOrbital;}
         ifile2.close();
    }
    if (spinstate==1){
         std::ifstream ifile2("lastActiveOrbital_down.csv");
         if (ifile2){ ifile2 >> lastActiveOrbital;}
         ifile2.close();
    }
    assignment_spin.head(lastActiveOrbital) = Eigen::VectorXi::Zero(lastActiveOrbital);
    spinstate++;
  };
  // TODO: Allow the splitting of core and non-core orbitals.
  orbitalController->updateOrbitals(coefficients, _supersystem->getActiveOrbitalController<SCFMode>()->getEigenvalues());
  outfile1.close();
  outfile2.close();
  return assignment;
}
template class SPADEAlgorithm<Options::SCF_MODES::RESTRICTED>;
template class SPADEAlgorithm<Options::SCF_MODES::UNRESTRICTED>;
} /* namespace Serenity */
```

**Figure S1.** Modified code of the Serenity 1.4 package SPADEAlgorithm.cpp. The modified parts are highlighted in green.

```
+system                                    +system
    name Au13_CO                               name Au12
    geometry Au13_CO.xyz                       geometry Au12.xyz
    method dft                                 method dft
    +dft                                       +dft
        functional LDA                             functional LDA
    -dft                                       -dft
    +basis                                     +basis
        label DEF2-SVP                             label DEF2-SVP
    -basis                                     -basis
    charge 0                                   charge 0
    spin 1                                     spin 0
-system                                    -system

+system                                    +task SCF
    name Au_CO                                 system Au13_CO
    geometry Au_CO.xyz                     -task
    method dft
    +dft                                   +task SPLIT
        functional PBE0                        act Au13_CO
    -dft                                       env Au_CO
    +basis                                     env Au12
        label DEF2-SVP                         systemPartitioning SPADE
    -basis                                     printLevel DEBUGGING
    charge 0                               -task
    spin 1
-system                         page 1                                       page 2
```

**Figure S2.** Exemplary Serenity input file for conducting the SPADE with the modified SPADEAlgorithm.cpp code. Starting Serenity with this input writes out the singular value distribution to the files "sigma_orb_up.txt" and "sigma_orb_down.txt".

```python
import numpy as np
from scipy.optimize import curve_fit

def deriv_fermi(x, x0, beta, scale):
    numinator=scale*beta*np.exp(beta*(x-x0))
    denominator=(1.+np.exp(beta*(x-x0)))**1.5
    return numinator/denominator

opt_id=[]
for spin in ["up", "down"]:
    matrix=[]
    infile=open("sigma_orb_"+spin+".txt", "r")
    id_max=0
    value=0
    i=0

    x_data=[]
    y_data=[]
    for line in infile:
        x_data.append(float(line.strip().split()[0]))
        y_data.append(float(line.strip().split()[2]))
        if float(line.strip().split()[2]) > value:
            value=float(line.strip().split()[2])
            id_max=i
        i+=1
    infile.close()

    popt, pcov = curve_fit(deriv_fermi, x_data, y_data, p0=[id_max, 1., 1.])
    opt_id.append(float(popt[0]))

for spin in ["up", "down"]:
    outfile=open("lastActiveOrbital_"+spin+".csv", "w")
    if spin == "up": outfile.write(str(opt_id[0]+1)) #+1 because the modified serenity code just reads the number of active MO, not the last index, which starts with index 0.
    if spin == "down": outfile.write(str(opt_id[1]+1)) #see explanation above
    outfile.close()
```

**Figure S3.** The "curve_fit.py" script. This script reads the singular value distributions stored in "sigma_orb_up/down.txt" and fits this curve with the first derivative of the square root of the Fermi-type function. The peak position is then saved as float in the files "lastActiveOrbital_up/down.csv". By placing these output files in the directory where the serenity calculation is conducted, the system partitioning is manipulated with respect to the information defined in these files.

```python
import numpy as np
import os

tmp_input=input("start values: ")
start_val=int(tmp_input.strip().split()[0])
end_val=int(tmp_input.strip().split()[1])

lastActiveOrbitals=0
for spin in ["up","down"]:
    act_orb=[]
    lastActiveOrbital=0
    for i in range(start_val, end_val):
        filename="./"+str(i)+"/lastActiveOrbital_"+spin+".csv"
        if os.path.isfile(filename):
            infile=open(filename,"r")
            for line in infile:
                tmp=float(line.strip().split()[0])
                act_orb.append(tmp)
            infile.close()
            If tmp > lastActiveOrbital: lastActiveOrbital=tmp
            else: pass
    if lastActiveOrbital==0:
        print("THERE IS NOT A SINGLE LASTACTIVEORBITAL FILE")
        exit()
lastActiveOrbital=np.average(act_orb)
outfile = open("lastActiveOrbital_"+spin+".csv","w")
outfile.write(str(int(lastActiveOrbital)))
outfile.close()
```

**Figure S4.** The "calc_ac_of_peaks.py" script. This script requires that the files "lastActiveOrbital_up.csv" and "lastActiveOrbital_down.csv" are saved in different directories numbered with respect to the corresponding structure along the trajectory. The script reads iteratively all "lastActiveOrbital_up/down.txt" files and calculates the average value. The integer number of the average value is then written to new "lastActiveOrbital_up/down.txt" files. This is the basis for manipulating the system partitioning during a subsequent embedding calculation using the modified version of Serenity.

```
+system                                    +task SCF
    name Au13_CO                               system Au13_CO #optional: the previous
    geometry Au13_CO.xyz                                       calculated orbitals can be
    method dft                                                 read to skip this calculation
    +dft                                                       and to accelerate the whole
        functional LDA                                         procedure.
    -dft                                   -task
    +basis
        label DEF2-SVP                     +task SPLIT
    -basis                                     act Au13_CO
    charge 0                                   env Au_CO
    spin 1                                     env Au12
-system                                        systemPartitioning SPADE
                                               printLevel DEBUGGING
+system                                    -task
    name Au_CO
    geometry Au_CO.xyz                     +task FDE
    method dft                                 act Au_CO
    +dft                                       env Au12
        functional PBE0                        +EMB
    -dft                                           naddXCFunc PBE0
    +basis                                         embeddingMode FERMI
        label DEF2-SVP                         -EMB
    -basis                                     calculateEnvironmentEnergy True
    charge 0                                   printLevel DEBUGGING
    spin 1                                 -task
-system

+system
    name Au12
    geometry Au12.xyz
    method dft
    +dft
        functional LDA
    -dft
    +basis
        label DEF2-SVP
    -basis
    charge 0
    spin 0
-system                         page 1                                              page 2
```

**Figure S5.** Exemplary Serenity input file for conducting the final embedding calculation with the modified SPADEAlgorithm.cpp code. Starting Serenity with this input and the previously "lastActiveOrbital_up/down.csv"-files stored in the directories where Serenity is executed, reads in where the system partitioning occurs and conducts the separation of active and environmental systems accordingly.

## 2. Definition of the Test System

Figure S6 shows the CO@Au$_{13}$ test system used in this study. The minimum structure used as a template for all calculations is given below in Table S1.

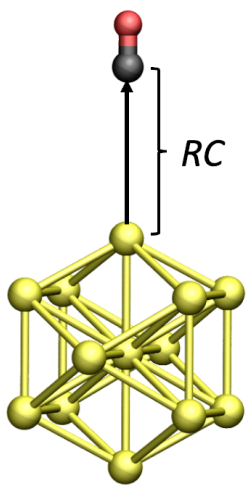

**Figure S6.** Illustration of the test system CO@Au$_{13}$. The reaction coordinate $RC$ is defined by the vector pointing from the top Au atom to the C atom of the substrate.

| 15 | | | |
|---|---|---|---|
| C  | 3.5456393166760583 | 0.6083152873033600  | 4.7395663983811070 |
| O  | 3.5456393166760583 | 0.6083152873033600  | 5.87487367795780700 |
| Au | 3.5456393166760583 | 0.6083152873033600  | 2.7395663983811067 |
| Au | 3.5456393166760583 | 0.6083152873033600  | 7.474353716937067e-07 |
| Au | 1.0812310140118795 | 1.0746189400920991  | 1.4128783326884087 |
| Au | 3.0793180446227852 | 3.0728191605219006  | 1.4129248011236335 |
| Au | 4.0119418581793770 | -1.856085882304913  | 1.4128740516959597 |
| Au | 6.0101354931901465 | 0.1419948119741670  | 1.4129207605189390 |
| Au | 1.6085409127884487 | -1.328784654606081  | 5.673546180810722e-06 |
| Au | 5.4828962177637140 | 2.5455709125561348  | -2.72215299647758e-06 |
| Au | 1.0812135215815302 | 1.0746195529897853  | -1.4128812237879000 |
| Au | 4.0119424038699960 | -1.856103422579450  | -1.4128741879203162 |
| Au | 3.0793120535051948 | 3.0728310848040390  | -1.4129375215170563 |
| Au | 6.0101468807459850 | 0.1419884309802613  | -1.4129303290171256 |
| Au | 3.5456334264436937 | 0.6083087693573925  | -2.7395995321356534 |

**Tabel S1.** Coordinates of the probe system CO@Au$_{13}$ at 2.0 Å used as a template for all calculations.

### 3. Conventional Orbital Localizations

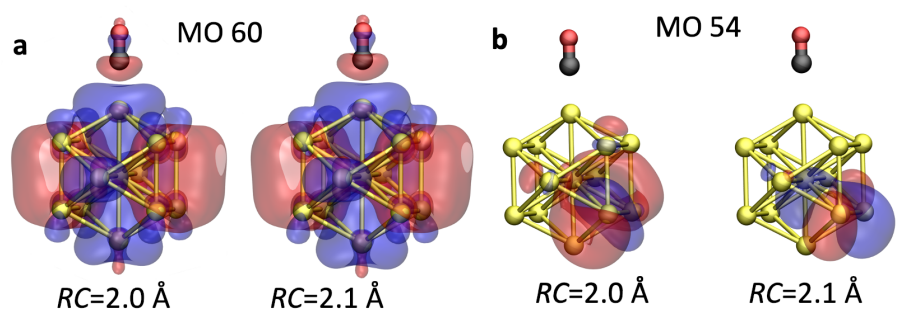

**Figure S7.** MOs after orbital localization using IBO. a) highlights that the atom-orbital-like characteristics are not always achieved. The image shows the MO with index 60 for a reaction coordinate $RC$=2.0 Å and $RC$=2.1 Å. b) shows that the localization might converge to different electronic minima for slightly different structures. Shown are the MO with index 54. Note, orbitals with different indices have also been compared here for better agreement, but none were found. The isosurfaces highlighted in blue and red correspond to isovalues of $\pm 0.0025$, respectively.

## 4. Singular Value Distributions

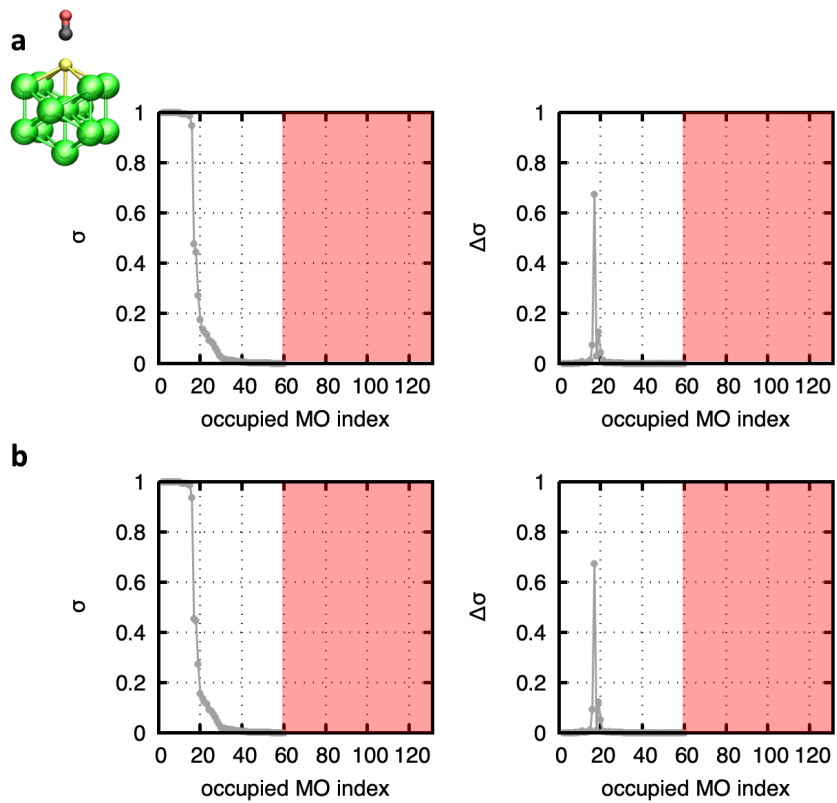

**Figure S8.** Plots of the singular value $\sigma_i$ distribution (left panel) and the difference of consecutive MOs with respect to the MO index ($\Delta\sigma_i = \sigma_i - \sigma_{i-1}$) for the CO@Au-in-Au$_{12}$ nanocluster (right panel). a) shows the distribution of the alpha and b) of the beta MOs. The red areas are not subject to the singular value decomposition procedure.

## 5. Error Compensation

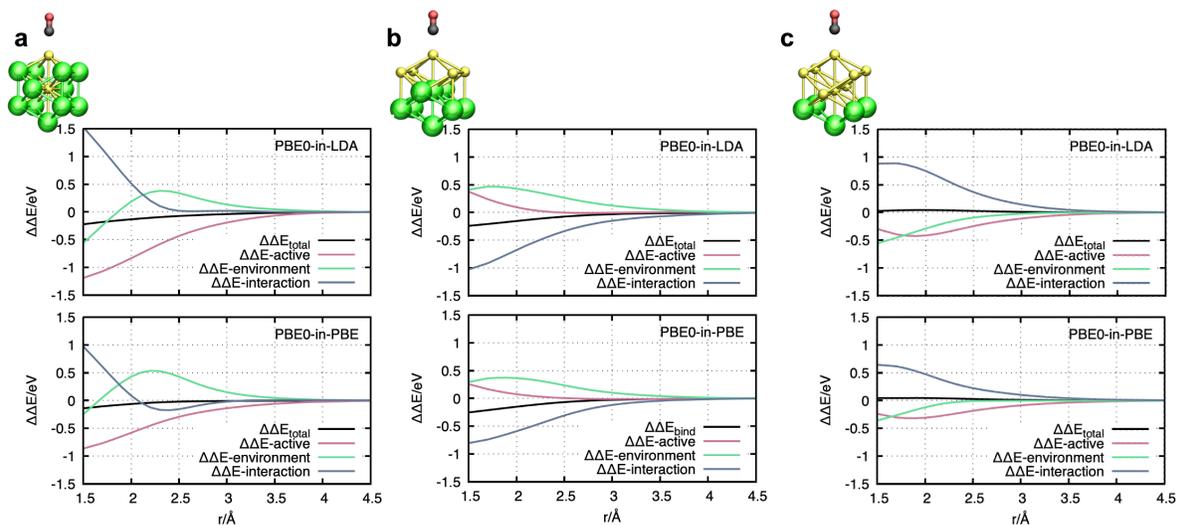

**Figure S9.** Individual error contributions for different systems with different active atom configurations. The black curves represent the total error given by the accumulation of individual contributions. The top diagrams show the calculated errors for treating the environment with the LDA functional, while the bottom diagrams show the environment with the PBE functional. The non-additive exchange interactions are described using PBE0. a) shows the errors for the cluster partitioning CO@$Au_2$-$Au_{11}$, b) for CO@$Au_5$-$Au_8$ and c) for CO@$Au_8$-$Au_5$.

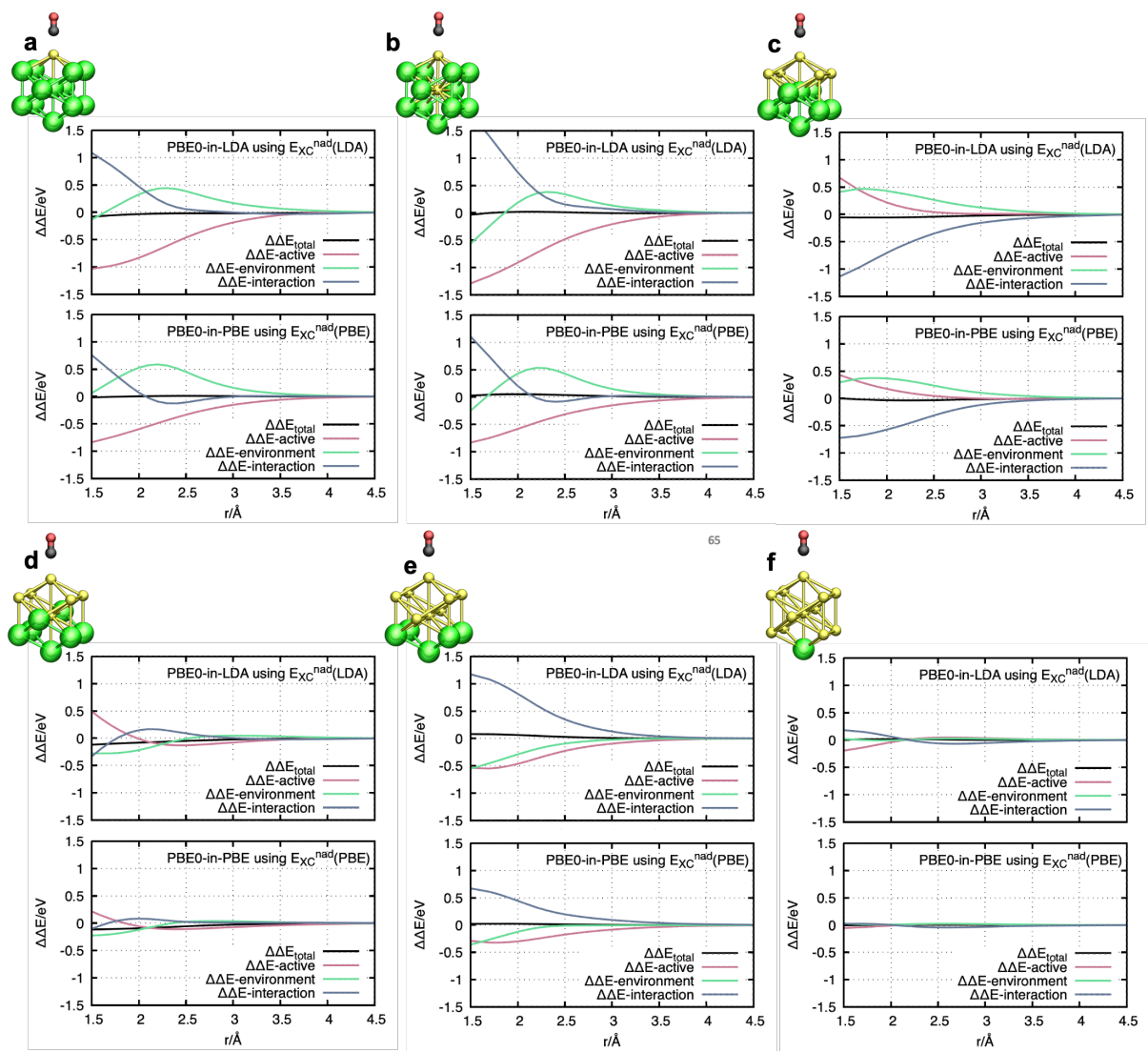

**Figure S10.** Individual error contributions for different systems with different active atom configurations. The black curves represent the total error given by the accumulation of individual contributions. The top diagrams show the calculated errors for treating the environment and the non-additive exchange interaction with the LDA functional, while the bottom diagrams show the environment and non-additive exchange interactions with the PBE functional. a) shows the errors for the cluster partitioning CO@Au-Au$_{12}$, b) for CO@Au$_2$-Au$_{11}$, c) for CO@Au$_5$-Au$_8$, d) for CO@Au$_6$-Au$_7$, e) for CO@Au$_8$-Au$_5$ and f) for CO@Au$_{12}$-Au